# Phase transformation in two-dimensional covalent organic frameworks under compressive loading

Jin Zhang

Shenzhen Graduate School, Harbin Institute of Technology, Shenzhen 518055, China

**Abstract:** As a new class of two-dimensional (2D) materials, 2D covalent organic frameworks (COFs) are proven to possess remarkable electronic and magnetic properties. However, their mechanical behaviours remain almost unexplored. In this work, taking the recently synthesised dimethylmethylene-bridged triphenylamine (DTPA) sheet as an example, we investigate the mechanical behaviours of 2D COFs based on molecular dynamics simulations together with density functional theory calculations. A novel phase transformation is observed in DTPA sheets when a relatively large in-plane compressive strain is applied to them. Specifically, the crystal structures of the transformed phases are topographically different when the compressive loading is applied in different directions. The compression-induced phase transformation in DTPA sheets is attributed to the buckling of their kagome lattice structures and is found to have significant impacts on their material properties. After the phase transformation, Young's modulus, band gap and thermal conductivity of DTPA sheets are greatly reduced and become strongly anisotropic. Moreover, a large in-plane negative Poisson's ratio is found in the transformed phases of DTPA sheets. It is expected that the results of the compression-induced phase transformation and its influence on the material properties observed in the present DTPA sheets can be further extended to other 2D COFs, since most 2D COFs possess a similar kagome lattice structure.

**Keywords:** Two-dimensional material; Covalent organic framework; Mechanical behaviour; Phase transformation

E-mail address: jinzhang@hit.edu.cn.

# 1. Introduction

Since the successfully synthesis of one atom thick graphene in 2004 [1], two-dimensional (2D) materials such as hexagonal boron nitride, transition metal-dichalcogenide, silica bilayer, black phosphorus, silicone and borophene have been the subject of intense research in the past decade [2-7]. It is found that these 2D materials possess unique properties significantly different to their three-dimensional bulk counterparts, which make them potential materials for application in next-generation nanoelectronics and nanodevices [8-12]. Inspired by the remarkable properties discovered in 2D materials together with their potential applications in various engineering practices, researchers recently have been trying to fabricate some other 2D materials at the nanoscale. Among these novel 2D materials, 2D covalent organic frameworks (COFs) have received special attention due to the rapid development of their synthesis method and also the high abundance of their molecular structures [13-18]. 2D COFs comprise a series of materials that are based on the atomically precise organization of organic precursor molecules into 2D crystalline structures connected by strong covalent bonds. In 2005, Yaghi and co-workers [19] reported the first 2D COF materials, which have a hexagonal structure. In addition to hexagonal COFs, COFs possessing other structures such as tetragonal and triangular COFs were also reported since then [20-22].

As a novel 2D materials with porous crystalline structures constructed by strong covalent bonds in a periodic arrangement entirely from light elements (i.e., H, B, C, N and O), 2D COFs are reported to possess numerous properties superior to the conventional 2D materials such as graphene, hexagonal boron nitride and so on [23-26]. For example, 2D COFs possess a rigid 2D structure, small mass density, high thermal stability, small band gap and permanent porosity with relatively large surface. The remarkable properties observed in 2D COFs render them appealing for the use in energy conversion and storage, gas storage and separation, catalysis, chemo-sensing, organic electronics [23-26]. It is well recognised that during the processes of synthesis and application, 2D COFs are usually subjected to various external mechanical stimuli [25]. However, not enough attention has been paid to the mechanics of 2D COFs. Previous studies on various other 2D materials indicate that the mechanical loading can significantly affect their physical and chemical properties [27-29]. Thus, understanding the mechanical properties of 2D COFs and the dependence of their physical and chemical properties on external mechanical loads plays a key role in the success of their future engineering applications.

Motivated by these ideas, by taking the dimethylmethylene-bridged triphenylamine (DTPA) sheet as an example, in the present study we study the mechanical behaviours of 2D COFs that are under the in-plane compression based on molecular dynamics (MD) simulations. A novel phase transformation is observed in DTPA sheets when the applied compressive strain is relatively large. Our MD simulations and density functional theory (DFT) calculations also reveal that the compression-induce phase transformation has significantly effects on the material properties such as Young's modulus, Poisson's ratio, band gap and thermal conductivity of DTPA sheets.

**2. Simulation models and simulation methods**

As we mentioned above, COFs are porous network polymers which are usually constructed by cores and linkers. To date, 2D COFs with various geometries of porosity have been reported, among which 2D COFs with the hexagonal porosity are the most common COFs synthesised in experiments [13-17]. Thus, in the present study we put our focus on these hexagonal COFs. Generally, hexagonal COFs have two different topological diagrams, i.e., the ($C_3$ + $C_2$) and ($C_3$ + $C_3$) schemes shown in Fig. 1a, which can be obtained through different synthesis methods. Specifically, as shown in Fig. 1a, ($C_3$ + $C_2$) hexagonal COFs can be obtained through the reaction between the linear building blocks with the $C_2$ symmetry (linkers) and the triangular building blocks with the $C_3$ symmetry (cores). The first reported COF material (i.e., COF-5 shown in Fig. 1b) is a typical COF possessing this topology, which is synthesised by the reactions of phenyl diboronic acid ($C_2$ symmetry) and hexahydroxytriphenylene ($C_3$ symmetry) [19]. The ($C_3$ + $C_3$) hexagonal COFs can be achieved through the reaction between two kinds of triangular building blocks both with the $C_3$ symmetry. The recently synthesised dimethylmethylene-bridged triphenylamine (DTPA) sheet [30] shown in Fig. 1c is a typical ($C_3$ + $C_3$) COF material. Comparing ($C_3$ + $C_2$) COFs with ($C_3$ + $C_2$) COFs, we find that although ($C_3$ + $C_2$) and ($C_3$ + $C_3$) COFs have different molecular structures, they both belong to the kagome lattice structure. Considering the fact that ($C_3$ + $C_2$) and ($C_3$ + $C_3$) COFs have the same kagome lattice structure, in the present study on the mechanical behaviours of 2D COFs we only consider the DTPA sheet, a typical ($C_3$ + $C_3$) COF to simplify our analysis without losing generality. Moreover, it is noted in the previous experimental study that DTPA sheets do not occur naturally and only can be synthesised on the metal surface, e.g., Ag(111) surface [30]. Under this circumstance, unless otherwise specified DTPA sheets suspended on Ag(111) shown in Fig. 2a will be considered in the present work.

In this work, the mechanical behaviours of DTPA sheets were investigated by MD simulations, which were performed by using the open-resource software LAMMPS with no periodic boundary conditions in all directions [31]. In our MD simulations, the C-C and C-H interactions in DTPA sheets were described by the intermolecular reactive empirical bond order (AIREBO) potential [32], while the C-N interaction was represented by the Tersoff potential [33].

Here, the values of parameters in the AIREBO and Tersoff potentials were, respectively, taken from Refs. [34, 35]. The atomic interactions in the Ag substrate were described by the embedded atom potential [36]. Meanwhile, the non-bonded van der Waals (vdW) interactions between the suspended DTPA sheet and the Ag substrate were described by the Lennard-Jones (LJ) atomic pair potential [37]. The values of LJ parameters utilized in the present simulations were obtained from the mixing rules [38] together with the parameter values taken from Refs. [39, 40]. To investigate the mechanical behaviours of DTPA sheets, the following procedure was employed in our MD simulations. First, the DTPA sheet structures that were initially built based on the lattice constant of graphene sheets were relaxed to the local minimum energy state using the conjugate gradient algorithm. Second, the obtained DTPA sheets were relaxed for 20 ps within the NVT ensemble (constant number of particles, volume and temperature) to reach their equilibrium state. Here, unless otherwise specified, the simulation temperature was set at 1 K to reduce the thermally induced fluctuation of atoms. Moreover, in the present MD simulations the positions and velocities of atoms were updated by the velocity Verlet algorithm with a time step of 0.5 fs. Last, the DTPA sheets were quasi-statically compressed along $x$ or $y$ direction by fixing one end of the DTPA sheet and simultaneously pushing the opposite end with a relatively low strain rate of 0.001 ps$^{-1}$ (see Fig. 2a).

## 3. Results and discussion

In the present study, we considered the DTPA sheets with a size of 20 nm×20 nm. In Fig. 2b we show the stress-strain relationship of these DTPA sheets when they are compressed along $x$ and $y$ directions, respectively. Here, the strain is defined as $\varepsilon = (L-L_0)/L_0$ with $L$ and $L_0$ being lengths of DTPA sheets after and before the deformation, respectively. Meanwhile, the stress $\sigma$ is taken as the arithmetic mean of the local stresses on all atoms [41]. We firstly focus on DTPA sheets compressed in the *x* direction. We can see from Fig. 2b that, when the applied compressive strain $\varepsilon$ applied in the *x* direction is smaller than 3.5%, the stress $\sigma$ of the DTPA sheet increases linearly as the strain $\varepsilon$ grows, which corresponds to a linear elastic deformation during this loading process. Following the well-known Hooke's law, the Young's modulus of the DTPA sheet thus can be identified by the slope of the $\sigma$-$\varepsilon$ curve in this small deformation region. The obtained Young's modulus of the DTPA sheet is 418.7 GPa (see Tab. 1), which is greatly smaller than ~915 GPa of its graphene sheet counterpart (see Fig. S1 shown in the Supporting Information). The smaller Young's modulus observed in the DTPA sheets can be attributed to their unique porous structures [42]. When the compressive strain applied in the *x* direction goes beyond 3.5%, although $\sigma$ and $\varepsilon$ also show a nearly linear relationship, the slope of the $\sigma$-$\varepsilon$ curve drops to 124.6 GPa. In other words, the Young's modulus of the DTPA sheet can abruptly drop from 418.7 GPa to 124.6 GPa when the compressive strain applied in the *x* direction is larger than 3.5%. The instant change of the elastic behaviours in the DTPA sheet at the strain of -3.5% implies that a phase transformation may occur in the DTPA sheet subjected to a relatively large compressive strain. In Fig. 2c we show the atomistic structures of the DTPA sheets under the compressive strain smaller and larger than 3.5%. A topological difference is observed between these two atomistic structures, indicating that a phase transformation indeed occurs in the DTPA sheets. Specifically, from Fig. 2c we see that when the compressive strain is smaller than 3.5%, the DTPA sheet keeps the atomistic structure similar to its counterpart in the absence of the external loading, which belongs to the P6/mmm space group. Here, we note this crystal structure of DTPA sheets as phase I. After a compressive strain larger than 3.5% is applied, we find that some linkers (C-C bonds connecting two neighbouring DTPA flakes) between some neighbouring cores (DTPA flakes) in DTPA sheets bend, causing the rotations of some cores in DTPA sheets. These deformations result in a new crystal structure of DTPA sheets (noted as phase II), which possesses the pgg space group symmetry.

When shifting the compressive strain from the $x$ direction to the $y$ direction, in Fig. 2b we show the stress $\sigma$ against the strain $\varepsilon$ in the $y$ direction. The obtained $\sigma$-$\varepsilon$ curve is found to be extremely similar to that obtained above in the DTPA sheets compressed in the $x$ direction. The similarity between these two cases indicates that a similar phase transformation occurs in the DTPA sheets subjected to a compressive strain in the $y$ direction. As for the DTPA sheets that are compressed in the $y$ direction, their phase transformation is found to happen at a critical strain of -4.8%, which is smaller than -3.5% observed in their counterparts compressed in the $x$ direction. Moreover, in this case the Young's modulus obtained before the phase transformation is 421.6 GPa, which is extremely close to the value 418.7 GPa obtained in the $x$ direction. The almost same Young's modulus observed in the $x$ and $y$ directions of DTPA sheets before the phase transformation indicates that the parent phase (i.e., phase I) of DTPA sheets possess an isotropic elastic behaviour. However, after the phase transformation, the Young's modulus of DTPA sheets compressed in the $y$ direction is 55.5 GPa, which is 55% smaller than the value obtained in the transformed phase of DTPA sheets compressed in the $x$ direction. The significantly different Young's moduli observed in DTPA sheets compressed in $y$ and $x$ directions indicate that, after the phase transformation the new crystal structure of DTPA sheets compressed in the $y$ direction should be different to that of DTPA sheets compressed in the $x$ direction, i.e., phase II. In Fig. 2c we show the phase transformation-induced new crystal structure of DTPA sheets compressed in the $y$ direction, which is noted as phase III here. From Fig. 2c we can see that phase III possesses the cmm space group symmetry, which is indeed topologically different from the previous phase II which holds a lattice with pgg space group symmetry.

From Fig. 2c we see that although the new phases, i.e., phase II and phase III of DTPA sheets after the phase transformation show topologically different crystal structures, they are both triggered by the bending of the linkers and the rotation of the triangular cores in the kagome structure of DTPA sheets. This fact indicates that the compression-induced phase transformation occurring in DTPA sheets may be attributed to their unique kagome structures. Inspired by this idea, to better understand the mechanism behind the compression-induced phase transformation observed in DTPA sheets, we conducted finite element (FE) simulations on a kagome lattice structure subjected to the similar uniaxial compression. Since in the kagome lattice structures the hinges connecting neighbouring "rigid" triangular units have relatively small bending stiffness, they are prone to buckle even when the kagome lattice structure is subjected to a relatively small external strain. The buckling deformations of hinges thus trigger the rotations of triangular units connected to them, which finally results in the new distorted kagome lattice structures after the buckling instability as shown in Fig. 2c. The distorted structures of kagome lattice structures after the buckling instability obtained in FE simulations are found to extremely resemble the new phases (phase II and phase III) of DTPA sheets after the phase transformation obtained in MD simulations. This fact indicates that the compression-induced phase transformation observed in DTPA sheets can be attributed to the buckling instability of their unique kagome lattice structures. As we mentioned above, most 2D COFs have the similar kagome lattice structure. Thus, it is reasonably expected that the similar phase transformation phenomenon observed in the present DTPA sheets also exist in other 2D COFs when they are subjected to the in-plane compressive loading.

In the above study we have considered the DTPA sheets supported on the metal substrate. We also extended our simulations to the free-standing DTPA sheets. When the freestanding DTPA sheet with single layer is compressed by the in-plane loading, no phase transformation is observed. This is because in this case the global buckling instability of DTPA sheets occurs prior to the phase transformation, which results in the out-of-plane deformation of the entire DTPA structure (see Fig. S2 in the Supporting Information). As for the substrate supported DTPA sheets considered above, their transverse deformations can be greatly constrained by the vdW forces between the substrate and suspended DTPA sheets, which will significantly postpone the buckling of entire DTPA sheets. As a result, in the substrate supported DTPA sheets the phase transformation can happen prior to the buckling instability. When we gradually increase the layer number of the freestanding DTPA sheets to seven, we find the reappearance of the phase transformation (see Fig. S3 in the Supporting Information). According to the Euler buckling theory [43], it is known that the critical buckling strain of a structure is proportional to its thickness. Thus, by increasing the layer number of freestanding DTPA sheets we can similarly postpone their global buckling instability and thus make the phase transformation occur prior to the global buckling instability.

From the above discussion, we see that the compression-induced phase transformation can exist in the substrate supported DTPA sheets and also the freestanding multi-layered DTPA sheets. Moreover, the phase transformation can affect the mechanical properties of DTPA sheets by greatly reducing their Young's modulus. As another important material parameter characterising the mechanical behaviours of DTPA sheets, we are also interested in the influence of the phase transformation on the Poisson's ratio of DTPA sheets. Generally, the Poisson's ratio of a material can be defined as $\nu = -\partial\varepsilon_t/\partial\varepsilon_a$, where $\varepsilon_t$ and $\varepsilon_a$ are strains in the loading and the corresponding perpendicular directions, respectively. Thus, to evaluate the Poisson's ratio of the DTPA sheets, in Fig. 2b we show the transverse strain against the axial strain of DTPA sheets when they are compressed along $x$ and $y$ directions, respectively. From Fig. 2b we can see that before the phase transformation the transverse strain grows as the applied axial strain increases, which indicates a positive value of the Poisson's ratio in the parent phase I of DTPA sheets. The obtained Poisson's ratio of DTPA sheets within phase I is around 0.34, which is almost independent with the direction of DTPA sheets. However, after the phase transformation occurs, the transverse strain begins to decrease with further increasing the axial strain, which corresponds to a negative Poisson's ratio in phase II and phase III of DTPA sheets. The obtained Poisson's ratios of DTPA sheets within phase II and phase III are, respectively, -0.18 and -0.29. Actually, a similar negative Poisson's ratio behaviour was also observed in the recent experiment study [44] on a macroscale distorted kagome lattice structure having the structure similar to phase II of DTPA sheets considered here. Moreover, in the experiment [44] the Poisson's ratio of the phase II-like distorted kagome lattice structure was reported to be around -0.2, which is close to -0.18 detected in the present MD calculations. After analysing the crystal structures of DTPA sheets after the phase transformation (see Fig. 2c), we can find that the negative Poisson's ratio in phase II and phase III of DTPA sheets can be attributed to the fact that in phase II and phase III the compressive loading will trigger the significant rotation deformations of triangular DTPA units, which will shorten the distance between two neighbouring triangular DTPA units (e.g., the distance between points $a$ and $b$ in Fig. 2c) and thus results in the shrink of DTPA sheets in their transverse direction. It is worthy that the negative in-plane Poisson's ratio intrinsically existing in DTPA sheets may render them much more appealing in applications such as sound and vibration absorption and design of tougher materials when compared to the conventional 2D materials.

In the above discussion we find that the mechanical properties of DTPA sheets before and after phase transformation are distinctly different. Since the phase transformation can topologically change the crystal structure of DTPA sheets, it is believed that, in addition to mechanical properties, some other material properties of DTPA sheets also can be significantly changed due to the occurrence of phase transformation. In what follows, we will give a brief discussion on the possible influence of phase transformation on the electronic and thermal properties of DTPA sheets. We firstly studied the electronic properties of DTPA sheets subjected to the in-plane compression. In doing this, DFT calculations were performed to the atomistic structures of DTPA sheets under different compressive strains, which can be extracted from the above MD simulations. Here, all DFT computations were implemented in the CASTEP package by using the generalized gradient approximation of the Perdew-Burke-Ernzerhof functional form [45]. In Fig. 3a we show the electronic band structure of DTPA sheets which are in the absence of external strains. A small band gap of 0.43 eV is found in Fig. 3a, which indicates the semiconducting characteristics of DTPA sheets. Moreover, the value of 0.43 eV obtained in the present calculation is found to be close to ~0.5 eV reported in the previous study [25]. When an in-plane compressive load is applied, we find in Fig. 3b that before the phase transformation the band gap of the DTPA sheets gradually decreases with increasing the applied strain. For instance, drops of 20% and 30% are found in the band gap of DTPA sheets when they are, respectively, subjected to compressive strains of -3% in the $x$ direction and -4% in the $y$ direction. Here, the compression-induced reduction in the band gap of DTPA sheets is probably due to the competition between near-band-edge states [46]. After the applied strain goes beyond the critical value characterising the phase transformation behaviours of DTPA sheets, although no significant changes are observed in their charge density (see Fig. 3c), a significant impact of the phase transformation on the band gap of DTPA sheets is detected as shown in Fig. 3b. Specifically, it is found that although the DTPA sheets within the transformed phase II and phase III retain the semiconducting characteristics of their parent phase I counterparts, the band gap of phase II and phase III is much larger than that of phase I. For example, as for DTPA sheets within phase I and under a compressive strain of -3% in the *x* direction, their band gap is 0.34 eV. When we slightly increase the applied compressive strain to -4%, the DTPA sheets transform to phase II accompanying with a drop of their band gap to only 0.075 eV. A similar abrupt drop of the band gap is also found in the phase I-to-phase III transformation when the DTPA sheets are compressed in the *y* direction. For instance, the band gap is 0.3 eV for the DTPA sheets within phase I and under a compressive strain of -4% in the *y* direction. The value of the band gap, however, significantly drops to 0.054 eV when the DTPA sheet transitions to phase III at the strain of -5%. Within transformed phase II and phase III, if we keep increasing the compressive strain the band gap will decrease correspondingly, which is similar to the findings detected in their phase I counterparts.

Finally, we studied the influence of the phase transformation on the thermal conductivity of DTPA sheets. In doing this, non-equilibrium molecular dynamics (NEMD) simulations [47] were performed to the equilibrium configurations of DTPA sheets with different crystal structures (phase I, phase II and phase III) achieved in the above MD simulations. In NEMD simulations, two ends of the DTPA sheets were used as the hot and cold reservoirs, respectively (see Fig. 4a). At each time step of NEMD simulations, a small amount of heat was added to the hot reservoir and removed from the cold reservoir, which was implemented by modifying the kinetic energies through a velocity rescaling procedure in both hot and cold reservoirs within the microcanonical ensemble. This treatment resulted in a constant heat flux $J$ across the middle region between hot and cold reservoirs and thus a temperature gradient $\partial T/\partial \alpha$ in the axial direction as shown in Fig. 4b, where $\alpha = x$ or $y$ represents the heat flux direction. Thus, the thermal conductivity $k$ can be achieved through the Fourier's law: $k = J/(A\partial T/\partial \alpha)$, where $A$ is the cross-sectional area of DTPA sheets. In calculation of the cross-sectional area the DTPA sheets were assumed to have a thickness same to that of their graphene counterparts, i.e., 3.4 Å [48]. Here, all NEMD simulations were conducted at room temperature (300 K).

After preforming aforementioned NEMD calculations to DTPA sheets, we obtained the thermal conductivity of their three phases. The results are graphically shown in Fig. 4c. From Fig. 4c we see that the thermal conductivity of the parent phase I of DTPA sheets is around 25 $Wm^{-1}K^{-1}$ and is almost independent with the direction of DTPA sheets, which indicates the isotropic thermal conductive behaviours of the parent phase I of DTPA sheets. The thermal conductivity of the parent phase I of DTPA sheets is found to be much smaller than the value of their graphene counterparts with the same geometric size (the thermal conductivity of graphene is predicted to be around 160 $Wm^{-1}K^{-1}$). The smaller thermal conductivity detected in DTPA sheets can be attributed to their porous structures and their hydrogenated edges, since the porosity of DTPA sheets will reduce the heat capacity [49, 50], while their hydrogen passivated edges will increase the phonon-defect scattering [51, 52]. After the phase transformation, the thermal conductivity of the transformed phases of DTPA sheets is much smaller than the value of their parent phase. For example, the thermal conductivities of the transformed phase II are 17.54 $Wm^{-1}K^{-1}$ and 11.24 $Wm^{-1}K^{-1}$ in $x$ and $y$ directions, which are, respectively, 30% and 55% smaller the value of its parent phase I counterpart. As for the transformed phase III, its thermal conductivities in $x$ and $y$ directions are 15.47 $Wm^{-1}K^{-1}$ and 7.24 $Wm^{-1}K^{-1}$, which are, respectively, 38% and 71% smaller than the value of its parent phase I counterpart. To better understand the observed significant influence of phase transformation on the thermal conductivity of DTPA sheets, in Fig. 4d we show the phonon density of states (PDOS) of DTPA sheets with different crystal structures. Here, the PDOS was calculated by performing the Fourier transform to the velocity autocorrelation function obtained in MD simulations. In Fig. 4d we find that there are two frequency peaks in PDOS, which locate around 46 THz and 87 THz. Here, the larger frequency corresponds to the vibration of C-H bonds, while the smaller frequency reflects the vibration of C-C and C-N bonds. Although, three phases of DTPA sheets almost have the same frequency, the amplitudes of two frequency peaks of transformed phase II and phase III are both smaller than those of parent phase I, which indicates more phonon scattering and thus smaller thermal conductivity in phase II and phase III of DTPA sheets. Obviously, the results predicted from PDOS are consistent with our NEMD calculations shown in Fig. 4c. It is worth mentioning that the flattening of phonon mode peaks observed in phase II and phase III can be attributed to the buckling deformations of some C-C bonds in them [49, 53, 54]. This fact also can be utilized to explain the anisotropic thermal conductive behaviours observed in transformed phase II and phase III of DTPA sheets. As shown in the inset of Fig. 4c, in the $y$ direction of phase II and phase III the heat flux need to pass through more buckled C-C bonds when compared to the case of the $x$ direction. As a result, more phonon scattering is expected to exist in the $y$ direction of phase II and phase III, which results in smaller thermal conductivity in the $y$ direction of the transformed phase II and phase III. This analysis result is in good agreement with the strongly anisotropic thermal conductivity of phase II and phase III observed in NEMD calculations (see Fig. 4c).

## 4. Conclusions

In this paper, the mechanical behaviours of DTPA sheets under the in-plane compression are investigated by MD and DFT simulations. A novel phase transformation is observed in DTPA sheets when the compressive strain is larger than a critical value. Specifically, the transformed phase of DTPA sheets is found to possess different crystal structures when they are compressed in different directions. The compression-induced phase transformation observed in DTPA sheets is attributed to the buckling instability of their unique kagome lattice structures. The phase transformation is found to significantly affect the material properties of DTPA sheets. Our results show that the Young's modulus, the band gap and the thermal conductivity of DTPA sheets are greatly reduced and become strongly anisotropic after the phase transformation. More importantly, a relatively large in-plane negative Poisson's ratio is found to intrinsically exist in the transformed phases of DTPA sheets. The results of compression-induced phase transformation and its influence on the material properties of observed in the present DTPA sheets can be further extended to other 2D COFs, since most 2D COFs possess the similar kagome lattice structure. It is expected that the significantly tunable material properties of 2D COFs due to their unique phase transformation may make them have more potential engineering applications when compared to the conventional 2D materials.

**Acknowledgements**

This work was supported by the National Natural Science Foundation of China (No. 11602074). The author also acknowledges the financial support from Harbin Institute of Technology (Shenzhen Graduate School) through the Scientific Research Starting Project for New Faculty.

**Table**

**Table 1.** Young's modulus and Poisson's ratio of the parent phase and the transformed phases of DTPA sheets

|  | Phase I | Phase II | Phase III |
|---|---|---|---|
| Young's modulus (GPa) | 418.7-421.6 | 124.6 | 55.5 |
| Poisson's ratio | 0.33-0.35 | -0.18 | -0.29 |

**Figures**

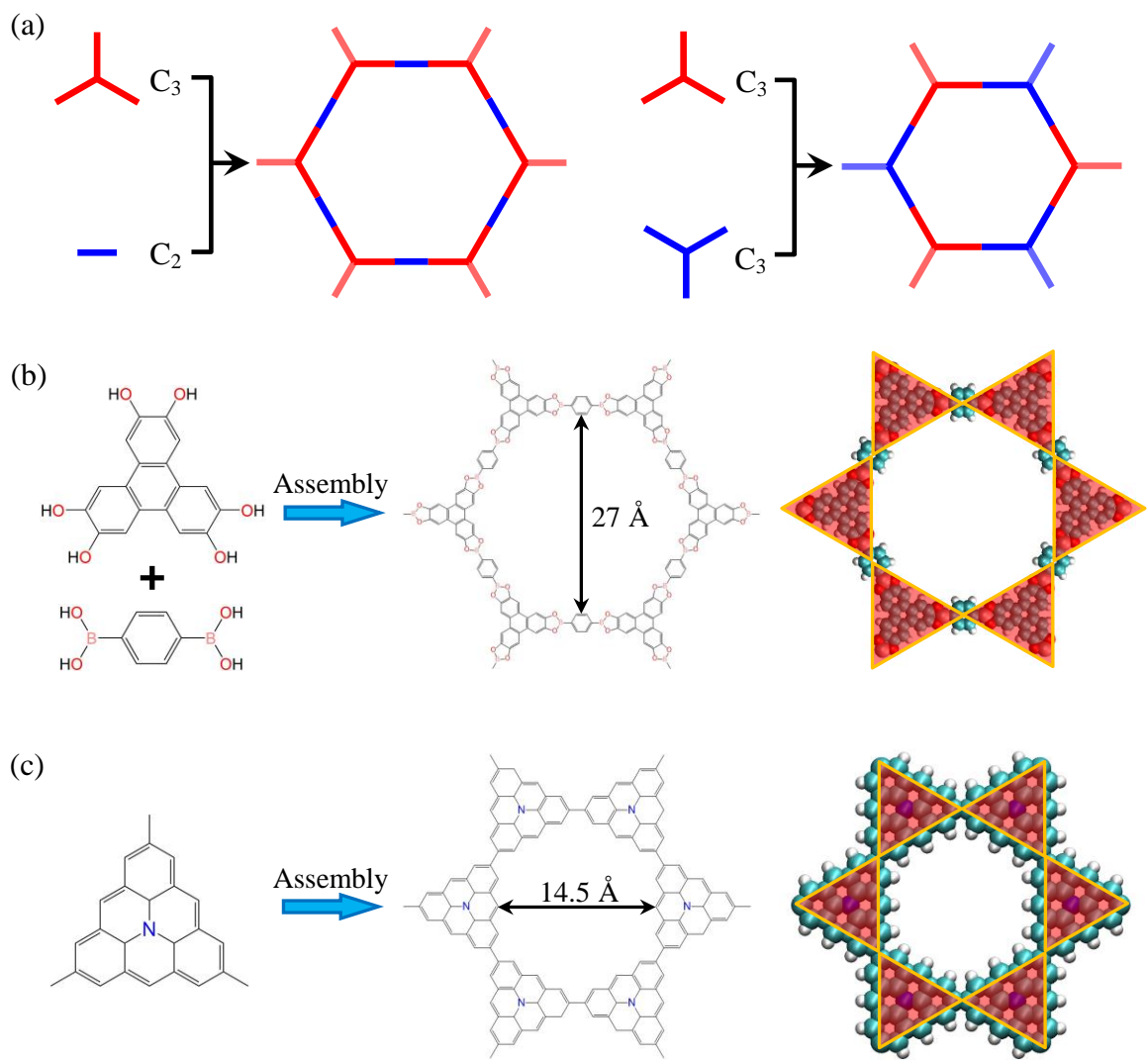

**Figure 1.** Structures of 2D COFs. (a) Two typical topology diagrams of 2D COFs. (b) Molecular structures of COF-5 and their building blocks. (c) Molecular structures of DTPA sheets and their building blocks. Both COF-5 and DTPA sheet have a kagome lattice structure.

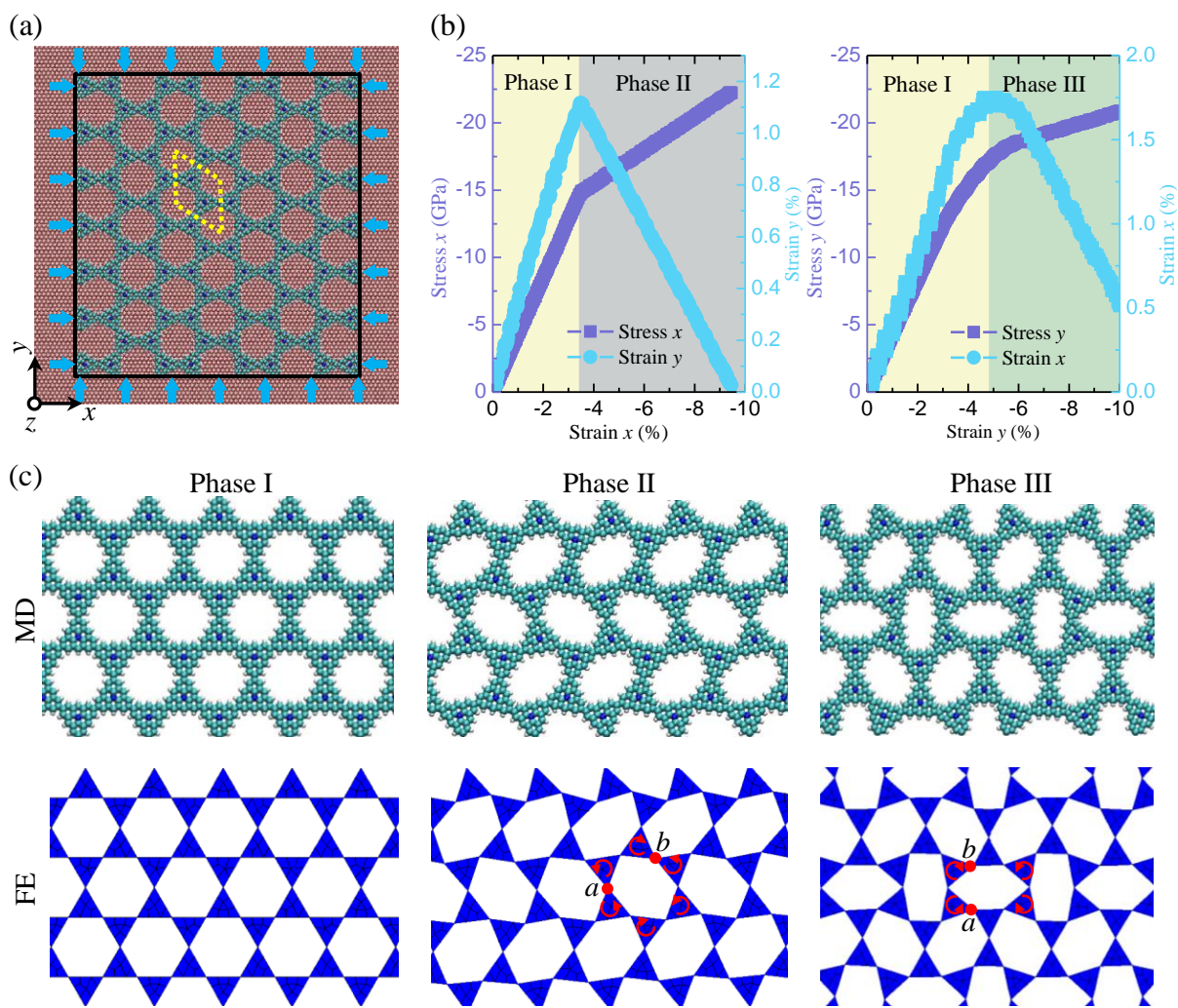

**Figure 2.** Mechanical behaviours of DTPA sheets subjected to the in-plane compression. (a) The schematic of a DTPA sheet grown on Ag(111). Here, the arrows denote the in-plane compression and the diamond represents the conventional unit cell. (b) The stress-strain relationship and the relationship between transverse strain and applied axial strain of DTPA sheets compressed in the *x* direction (left) and the *y* direction (right). (c) Crystal structures of the parent phase (phase I) and the transformed phases (phase II and phase III) of DTPA sheets. Structures of the transformed phases obtained in MD simulations (top) resemble the buckling modes of the corresponding kagome lattice structures of DTPA sheets obtained in FE simulations (bottom).

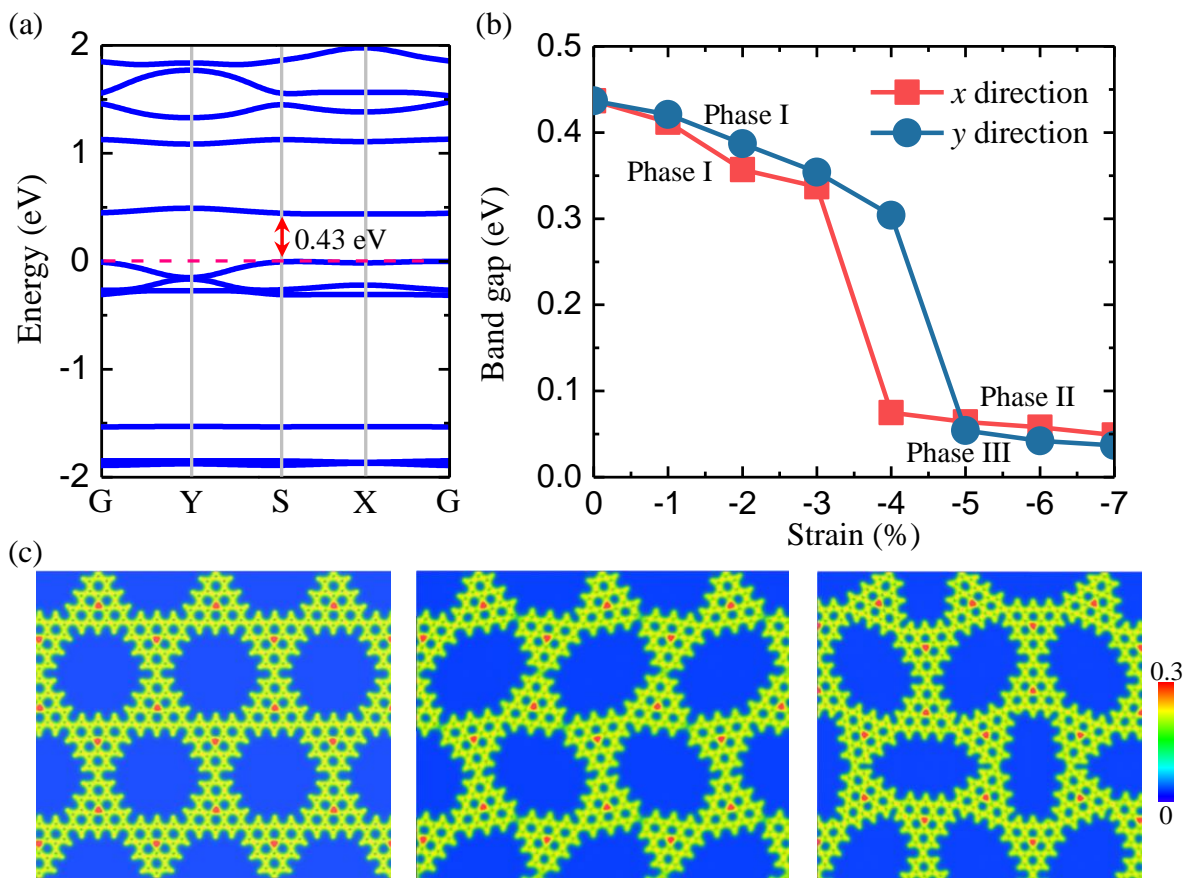

**Figure 3.** Electronic properties of the parent phase and the transformed phases of DTPA sheets. (a) Electronic band structures of DTPA sheets in the absence of external loading. (b) Band gaps of DTPA sheets subjected to different uniaxial compressive strains in the *x* and *y* directions, respectively. Here, an abrupt drop of the band gap occurs in the phase transformation process. (c) Valence electron densities of the parent phase and the transformed phases of DTPA sheets.

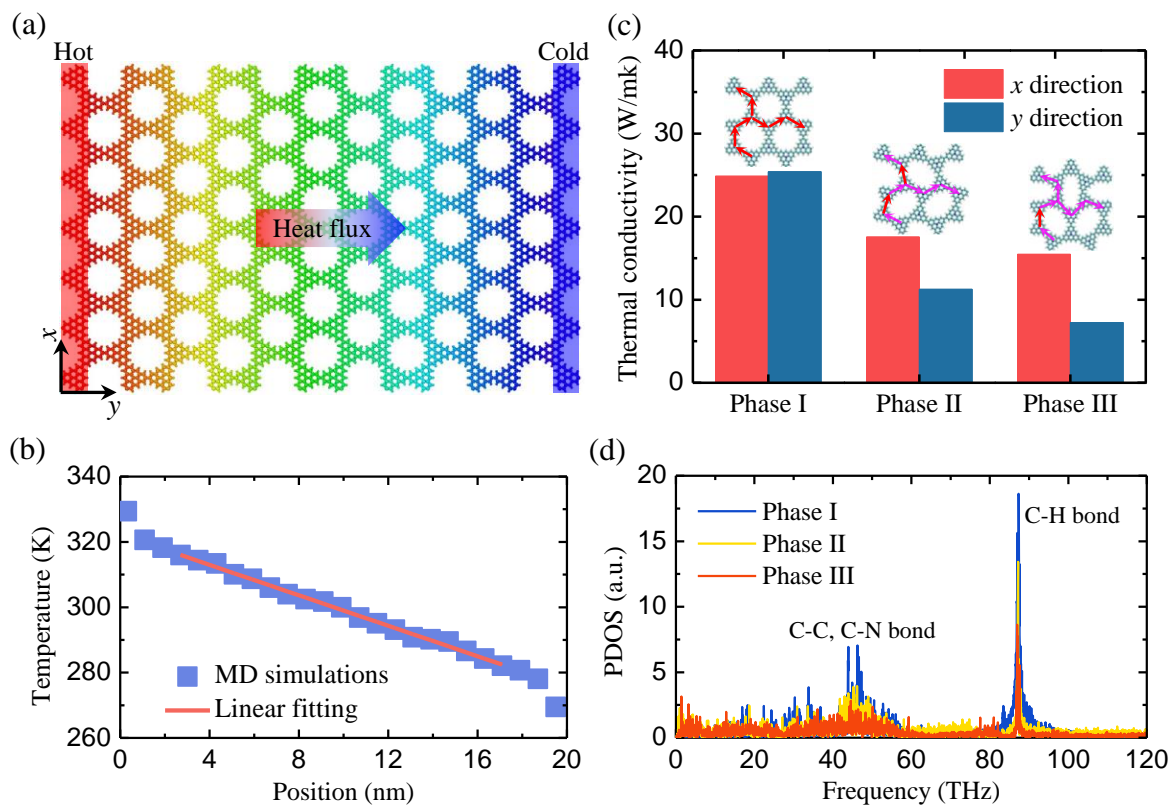

**Figure 4.** Thermal properties of the parent phase and the transformed phases of DTPA sheets. (a) NEMD simulation model for the heat transport in DTPA sheets. Here, the atoms are coloured according to their temperature. (b) Typical temperature distribution in the DTPA sheets along the heat flux direction. (c) The thermal conductivity of the parent phase and the transformed phases of the DTPA sheets. The arrows in the insets represent the heat flux channels in the DTPA sheets. (d) PDOSs of the parent phase and the transformed phases of the DTPA sheets.